\documentclass[aps,pra,twocolumn,
groupedaddress,superscriptaddress,
bibnotes,amsfonts,longbibliography,
citeautoscript,a4paper,
floatfix]{revtex4-2}

\usepackage{amsmath}
\usepackage{amssymb}
\usepackage{graphicx}
\usepackage{dcolumn}
\usepackage{bm}
\usepackage{orcidlink}
\hypersetup{colorlinks=true,allcolors=blue}
\usepackage{xcolor} 
\usepackage[normalem]{ulem}

\begin{document}

\preprint{APS/123-QED}

\title{Theoretical investigation of two-dimensional semiconductor\\ nanoribbons and nanoparticles for tailored light--matter interactions}

\author{Christian Nicolaisen Hansen\,\orcidlink{0009-0008-2350-7652}}
\email{nicolaisen@mci.sdu.dk}
\affiliation{POLIMA---Center for Polariton-driven Light-Matter Interactions, University of Southern Denmark, 5230 Odense M, Denmark}

\author{Line Jelver\,\orcidlink{0000-0001-5503-5604}}%
\affiliation{POLIMA---Center for Polariton-driven Light-Matter Interactions, University of Southern Denmark, 5230 Odense M, Denmark}

\author{Christos Tserkezis\,\orcidlink{0000-0002-2075-9036}}%
 \email{ct@mci.sdu.dk}
\affiliation{POLIMA---Center for Polariton-driven Light-Matter Interactions, University of Southern Denmark, 5230 Odense M, Denmark}
\affiliation{D-IAS---Danish Institute for Advanced Study,\\
University of Southern Denmark, 5230 Odense M, Denmark}


\date{\today}

\begin{abstract}
We explore theoretically the optical response of two-dimensional (2D) materials patterned at the nanoscale into either arrays of ribbons along a planar surface or spherical particles. Fourier-Floquet decomposition of the electromagnetic fields is used in order to obtain the reflectance, transmittance and absorbance of the nanoribbon array. The spherical particles are treated with Mie theory, with the boundary conditions modified to accommodate a surface conductivity at the interface. We consider the excitonic response of hexagonal boron nitride in the ultraviolet, and of the transition-metal dichalcogenide WS$_{2}$ in the visible. Unlike what is expected from graphene, where plasmons have proven very sensitive to geometry, the excitonic response of nanoribbon arrays is found to show weak tunability with the array parameters. To enhance tunability with the geometry of the system, or via hybridization with the substrate, nanospheres with a 2D semiconductor coating are demonstrated to provide a superior platform. Overall, we find that the localized nature of excitons in 2D semiconductors largely limits tunability with geometry, but their hybridization with other optical modes in the nanopatterning setup still holds promise for controllable light--matter interactions.
\end{abstract}

\maketitle

\section{Introduction}\label{sec1}
Two-dimensional (2D) materials have attracted much attention within nanophotonics since the emergence of graphene~\cite{Novoselov2004_Science306, Novoselov2005_Nature2005,Zhang2005_Nature2005}. These layered materials, where each of the atomically-thin layers are connected by van der Waals forces, have exhibited a wide diversity of new electronic and photonic phenomena~\cite{Zhang2016_2DMater3, Zeng2018_AdvElectronMater4, Cheng2019_AdvOptMater7}. For 2D materials exhibiting a band gap, external electromagnetic (EM) radiation can excite electrons from the valence band to the conduction band, leaving a hole in the valence band. The generated electron--hole pairs form charge-neutral quasiparticles known as excitons. Of special interest for this paper are the EM modes confined to the plane of the 2D semiconductor as the excitons couple to the external radiation. These modes are known as exciton polaritons, being one characteristic example among the wide range of possible polaritons~\cite{Basov2020_Nanophotonics10}. Here, inspired by the large collection of works investigating the plasmonic response of patterned graphene~\cite{Yan2012_NewJPhys14, Zhu2014_NanoLett14, Cox2014_NatureCommun5, Rasmussen2021_ACSPhotonics8}, we explore whether it is possible to tune the optical response of 2D semiconductors by patterning into nanoscale geometries. We explore the theory of such patterning in two different situations: hexagonal boron nitride (hBN), with a large band gap and ultraviolet (UV) activity, and tungsten disulfide (WS$_{2}$), a transition metal dichalcogenide (TMD)~\cite{Mak2010_PhysRevLett105, Zhu2011_PhysRevB84} with considerable excitonic response and a large oscillator strength in the visible regime.

The former material, hBN, has shown promise for the detection of biomolecules known as cyclic $\beta$-helical peptides~\cite{Nielsen1967_JPhysChem71, Bode1996_JPhysChem100, Fears2013_Langmuir29}. These molecules are important in the biophysics of the body and exhibit intense absorption at wavelengths between 200 and 240 nm. By using hBN as a material with a spectral response in the same UV range, one is able to investigate their chirality, namely the property of the molecule not being superposable onto its own mirror image. The atomically-flat surface of hBN, and its ability to support exciton polaritons in the UV, could therefore allow for the fabrication of optical sensors to detect these cyclic $\beta$-helical peptides. hBN is also a promising candidate for the fabrication of deep-UV optoelectronic devices and for solar-blind communication~\cite{Liu2018_Nanoscale10, Song2021_NatureCommun12, Wang2024_ACSOmega9, Xu2024_AdvMater36, Chen2025_AdvSci12}. On the other hand, the optical response of TMDs in general, and WS$_{2}$ in particular, in the visible regime, is of special interest for applications in optoelectronic devices. Compared to bulk semiconductors, these 2D materials have potential as ultrathin, flexible, and nearly transparent devices at low cost~\cite{Pospischil2014_NatNanotech9, Baugher2014_NatNanotech9, Ross2014_NatNanotech9, Lee2014_NatNanotech9, Jo2014_NanoLett14, Lopez2014_ACSNano8}.

Within nanophotonics, both hBN and WS\textsubscript{2} have been studied extensively due to their polaritonic response, with examples being phonon polaritons in the infrared (IR) for hBN~\cite{Caldwell2014_NatCommun5, Riveira2019_NanoLett19} and valley polaritons in TMDs~\cite{Sun2017_NatPhoton11}. The control of such polaritons has led to a multitude of applications, for instance within polaritonic transistors and lasing~\cite{Chakraborty2018_NanoLett18, Fan2024_ACSPhotonics11}, while at the same time
providing a fruitful platform for studies of fundamental
physics, e.g. in the context of Bose-Einstein
condensation, polariton blockade, polaritonic
simulators and optical nonlinearities~\cite{Basov2025_Nanophotonics14}. To further explore and advance such uses, tailoring the excitonic response of the materials might be beneficial. Here, we seek to understand whether experience from 2D-material plasmonics can be harnessed. The implementation of the methods described here enables the design and analysis of the response of nanoribbons and nanoparticles (NPs) including 2D semiconducting interfaces. This deeper understanding of the underlying material response will hopefully stimulate further advances in 2D-material nanopatterning, and facilitate the development of applications in polaritonic and nanophotonic devices. 

The paper is structured as follows; section~\ref{sec2} introduces the modeling of the optical conductivity of both hBN and WS\textsubscript{2} used throughout the paper. The optical conductivity will, for a given geometry, be implemented in the EM boundary conditions to introduce the 2D material into the system. Section~\ref{sec4} looks into methods for patterning 2D materials into an array of ribbons placed on a dielectric substrate. Finally, section~\ref{sec3} considers the patterning of the 2D material into spherical geometries; hollow shells and coatings around cores of a dielectric material. These geometries are modeled using methods that have been developed for single-layer graphene, but are now adapted for describing 2D semiconductors. While the investigation of excitonic phenomena in nanostructured 2D semiconductors is an active field~\cite{Nong2018_JPhysDApplPhys51, Goncalves2018_PhysRevB97, Qiang2025_Nanophotonics14}, to the best of our knowledge, there exist no similar works adapting these methods for 2D semiconductors.

\section{Optical conductivity of single-layer 2D materials}\label{sec2}
The optical properties of a single-layer 2D material can be contained in the angular frequency $\omega$-dependent optical conductivity $\sigma(\omega)$ at an infinitely thin interface. This method allows the 2D material to be introduced in any EM description through boundary conditions. Throughout this paper, we consider some interface at $\mathbf r_0$ between two media of electric and magnetic fields $\mathbf E_1,\mathbf H_1$ and $\mathbf E_2,\mathbf H_2$ with the unit vector $\hat{\mathbf n}$ normal to the interface. In the absence of surface currents, we have continuity of both the tangential electric and magnetic field across the interface. But if a 2D layer is placed along the interface, an optical surface conductivity is introduced, resulting in the tangential electric field $\mathbf E_\parallel$ at the interface $\mathcal S$ inducing a surface current given by Ohm's law as \mbox{$\mathbf K=\sigma(\omega) \mathbf E_\parallel |_{\mathcal S}$}. This causes a discontinuity of the tangential magnetic field components across the interface, giving conductive boundary conditions
\begin{align}
    \hat{\mathbf n}\times (\mathbf E_{2}-\mathbf E_{1})|_{\mathcal S}&=\mathbf 0,\label{eq:BCE}\\
    \hat{\mathbf n}\times (\mathbf H_{2}-\mathbf H_{1})|_{\mathcal S}&=\mathbf K.\label{eq:BCH}
\end{align}
These boundary conditions lead to a set of four linear equations that connect the two media with four unknowns as the electric and magnetic field amplitudes for two orthogonal polarizations each. A known optical conductivity of the 2D material $\sigma(\omega)$ is required in order to solve these equations for the EM fields of the system. This approach thus opens a fruitful route for nanophotonics and condensed-matter physics to join forces, enabling a direct implementation of the local, linear and isotropic optical conductivity of any 2D materials~\cite{UriaAlvarez2024_ComputPhysCommun295,Søndersted2024_PhysRevLett133} into traditional EM solvers. The methodology also assumes that the optical conductivity obtained for an extended monolayer of the 2D semiconductor remains the same once patterned into finite geometries. Changes in the band structure for atomistically small systems are outside the scope of this work. The conductivities given in this section might not be entirely accurate descriptions of the corresponding realistic materials, but are fully adequate for capturing the physics of the nanostructures explored in subsequent sections.

\subsection{Optical conductivity of hBN obtained from first-principle calculations}
\label{sec:hBN}
\begin{figure}[t]
\centerline{\includegraphics[width=\linewidth]{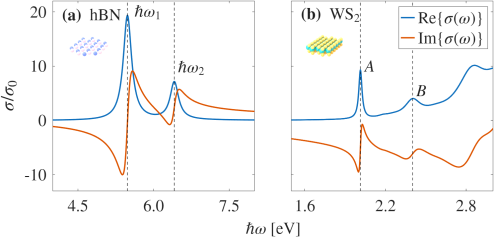}}
\caption{Optical conductivity of \textbf{(a)} monolayer hBN, obtained as discussed in Sec.~\ref{sec:hBN}, and \textbf{(b)} monolayer WS\textsubscript{2}, obtained as discussed in Sec.~\ref{sec:WS2}. Both optical conductivities are normalized to $\sigma_0 ={e^2}/{4\hbar }$ and given in terms of the real (imaginary) part as the blue (red) curve. The vertical dashed lines denote the main excitons in the two materials.
\label{optical_cond_both}}
\end{figure}

hBN is an insulating material, where boron and nitrogen atoms are arranged in a hexagonal crystal structure similar to graphene. In the IR, hBN exhibits phonon polaritons~\cite{Caldwell2014_NatCommun5}, but more interesting for this work is that hBN has shown strong excitonic resonances in the UV~\cite{Watanabe2004_NatMater3, Arnaud2006_PhysRevLett96}. Ferreira et. al.~\cite{Ferreira2019_JOptSocAmB} have determined the exciton energies from first-principles as $\hbar\omega_1=5.48\ \mathrm{eV}$ and $\hbar\omega_2=6.41\ \mathrm{eV}$ for hBN in the UV range. Inserting these exciton energies into the Elliot equation~\cite{Chaves2017_2DMater4}, we obtain the optical conductivity of single-layer hBN as 
\begin{align}
    \frac{\sigma^\mathrm{hBN}(\omega)}{\sigma_0 }=4\mathrm i\hbar \omega \sum _{n=1}^2 \frac{p_n }{\hbar \omega -\hbar \omega_n +\mathrm i \hbar \gamma_\mathrm E},\label{eq:cond:hBN}
\end{align}
where $\sigma_0 ={e^2}/{4\hbar }$ with $e>0$ being the elementary charge and $\hbar$ the reduced Planck constant. The summation is over the excitonic states \mbox{$n=\{1,2\}$}. A linewidth of \mbox{$\hbar\gamma_\mathrm{E}=0.1\ \mathrm{eV}$} is, unless otherwise stated, used throughout the text as an effective phenomenological broadening, with this choice motivated by the experimentally observed inhomogeneous broadening of the excitonic optical resonance in Ref.~\cite{Cassabois2022_PhysRevX12}. The parameters $p_1=0.088$ and $p_2=0.027$ correspond to the weight of each exciton energy as they are implemented in the Elliott equation. Figure~\ref{optical_cond_both}\textcolor{blue}{a} shows the real (imaginary) component of the optical conductivity given by the Elliott equation for hBN as the blue (red) curve in an energy window from $\hbar\omega=4\ \mathrm{eV}$ to $8\ \mathrm{eV}$. The exciton energies $\hbar\omega_1$ and $\hbar\omega_2$ are shown as vertical dashed lines. When excited by radiation matching these energies, the excitons carry charge across the atomically thin surface, leading to the non-zero optical conductivity.

\subsection{Optical conductivity of WS\textsubscript{2} obtained from experimental data}
\label{sec:WS2}
TMDs are a class of 2D semiconductors consisting of a layer of transition
metal atoms between two layers of chalcogen atoms, with the layers bonded by van der Waals forces. To better connect with experimental activities, instead of resorting to first-principles calculations we describe here WS\textsubscript{2} directly through experimental data. Either approach to obtaining the optical conductivity is applicable for the discussed methods. The optical conductivities of four different TMDs have been determined in Ref.~\cite{Li2014_PhysRevB90} from their experimentally measured reflectance spectra. We focus on WS\textsubscript{2} as a material with a strong excitonic response in the visible regime. The single layer of WS\textsubscript{2} is treated as a homogeneous medium with an effective thickness given as the interlayer spacing, $d=0.618\ \mathrm{nm}$, of bulk WS\textsubscript{2}. The relative permittivity is modeled as a linear combination of $N=9$ Lorentzian oscillators
\begin{align}
    \varepsilon^\mathrm {WS_2}(\omega) = \varepsilon_\infty+\sum _{k=1}^N\frac{f_k}{\hbar ^2\omega_k^2-\hbar ^2\omega^2-\mathrm i \hbar^2 \omega \gamma_k },\label{eq:epsWS2}
\end{align}
with $\hbar \omega_k$ being the resonance energy of each of the nine Lorentzian oscillators. The oscillator strength $f_k$ and linewidth $\gamma_k $ of each oscillator are fitted to the experimental data in the energy window from $\hbar \omega = 1.5\ \mathrm{eV}$ to $3.0\ \mathrm{eV}$, with the fitting parameters for WS\textsubscript{2} given in the supplementary material of Ref.~\cite{Li2014_PhysRevB90}. While a value of $\varepsilon_\infty$ for use in Eq.~\eqref{eq:epsWS2} is not given in the reference, here it has been chosen as $\varepsilon_\infty=10.68$ in order to align the value at the first local maximum of the real part of the optical conductivity closely to the curves given by both Ref.~\cite{Li2014_PhysRevB90} and Ref.~\cite{Hsu2019_AdvOptMat7}.  While small variations in the chosen value of $\varepsilon_\infty$ will slightly modify the magnitude of the optical response, such variations will not modify the conclusions of the following sections. The material response is now given as an optical sheet conductivity
\begin{align}
    \sigma^\mathrm {WS_2}(\omega) = -\mathrm i \varepsilon_0\omega [\varepsilon^\mathrm {WS_2}(\omega)-1]d,\label{eq:cond:WS2}
\end{align}
where the real and imaginary part are both shown in Fig.~\ref{optical_cond_both}\textcolor{blue}{b} normalized to $\sigma_0 ={e^2}/{4\hbar }$. The two lowest-energy peaks marked by vertical dashed lines in the optical conductivity spectrum correspond to the excitonic features associated with interband transitions at the $K$ ($K'$) points of the Brillouin zone, where WS\textsubscript{2} shows valence band maxima and conduction band minima, and where spin--orbit coupling causes a splitting of the valence band, leading to two distinct peaks~\cite{Zhang2014_PhysRevB89}. These peaks are denoted as the $A$ and $B$ lines of WS\textsubscript{2}. The broader response at higher energies is attributed to additional interband transitions.

\section{Excitons in periodic ribbons of 2D materials}\label{sec4}
\begin{figure}[t]
\centerline{\includegraphics[width=1\linewidth]{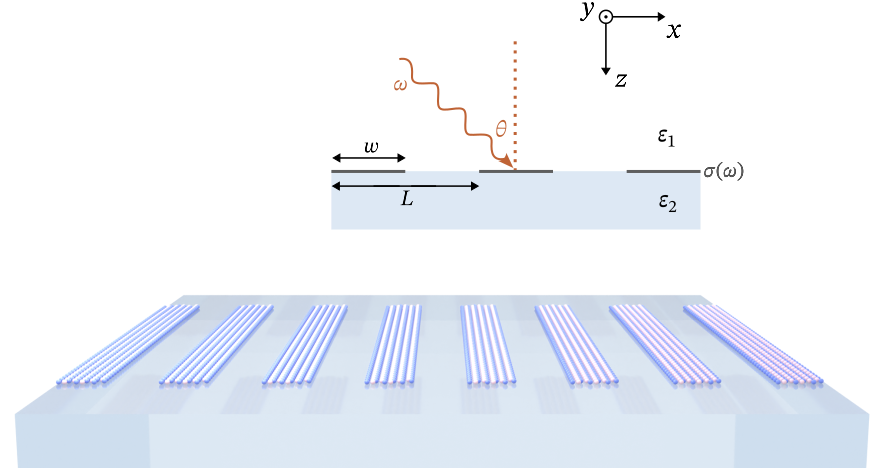}}
\caption{Schematic (upper) and illustration (lower) of the system described in Sec. \ref{sec4} consisting of nanoribbons of a 2D semiconductor of optical conductivity $\sigma(\omega)$ that are infinitely long in the $y$ direction, with a finite width $w$ and a periodicity $L$ along $x$. The interface of nanoribbons is placed between dielectrics $\varepsilon_1$ and $\varepsilon_2$, and it is illuminated by plane waves of angular frequency $\omega$ incident at an angle $\theta$ onto the interface. \label{fig:ribbon}}
\end{figure}

\begin{figure*}[ht]
\centerline{\includegraphics[width=\linewidth]{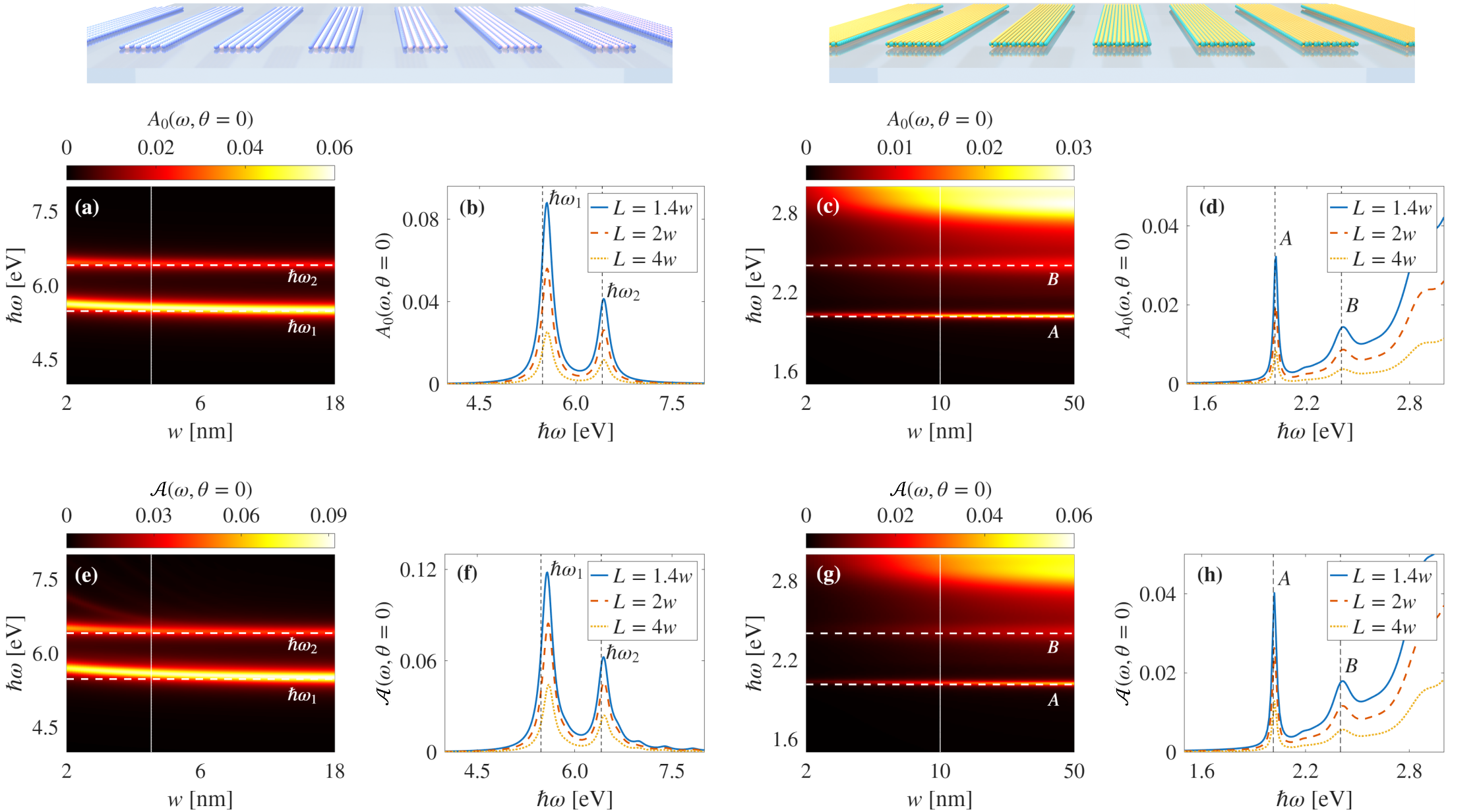}}
\caption{Absorbance of nanoribbons of a 2D semiconductor encapsulated by a dielectric $\varepsilon_1 = 3$ above, and a dielectric $\varepsilon_2 = 4$ below, with radiation at normal incidence $\theta=0$, using the edge condition method (panels \textbf{a}-\textbf{d}), or the Fourier series method (panels \textbf{e}-\textbf{h}). \textbf{(a)} Absorbance of hBN nanoribbons as a function of ribbon width $w$. The excitonic resonances of hBN are given as the horizontal dashed white lines. \textbf{(b)} Absorbance of hBN nanoribbons of width $w=4\ \mathrm{nm}$ for different periodicities $L$. \textbf{(c)} Absorbance of WS\textsubscript{2} nanoribbons as a function of ribbon width. The $A$ and $B$ peaks of WS\textsubscript{2} are given as the horizontal dashed white lines. \textbf{(d)} Absorbance of WS\textsubscript{2} nanoribbons computed using the edge condition method at $w=10\ \mathrm{nm}$ for different periodicities.
\textbf{(e)}-\textbf{(h)} Same as panels \textbf{(a)}-\textbf{(d)}, respectively, using the Fourier series method.}
\label{fig:nanoribbons}
\end{figure*}

In this section, we consider an array of ribbons along a planar interface as a well-established geometry for investigating the interaction of light with 2D materials. While having been proven challenging, the preparation of nanoscale TMD ribbons has previously been reported~\cite{Chen2017_NatureCommun8, Chowdhury2020_ChemRev120}. We focus on two tested methods~\cite{Goncalves2016_PhysRevB94, Goncalves2016} to analytically model such ribbons, that both assume an infinite array, as periodicity eases the calculation: an edge condition method and a Fourier series approach. For both methods, the optical properties of the single-layer 2D material are included as an optical conductivity. We consider a transverse magnetic (TM)-polarized wave of wave vector $\mathbf k_1=k_x\hat{\mathbf x}+k_z\hat{\mathbf z}$ in the \mbox{$xz$ plane}, initially in a medium $\varepsilon_1$ impinging at an angle $\theta$ onto a substrate $\varepsilon_2$, with the interface along $z=0$ as shown in Fig.~\ref{fig:ribbon}. This interface $z=0$ is patterned with ribbons of width $w$ of the 2D material to form a one-dimensional grating with a periodicity $L=2w$ along $x$. For this TM polarization, the EM fields incident on this grating are given as
\begin{align}
    \mathbf B^\mathrm i (\mathbf r,t)&=\hat{\mathbf y}B_0^\mathrm i \mathrm e^{\mathrm i (\mathbf k_1\cdot \mathbf r-\omega t)},\\
    \mathbf E^\mathrm i (\mathbf r,t)&=(\hat{\mathbf x}E_{0x}^\mathrm i +\hat{\mathbf z}E_{0z}^\mathrm i )\mathrm e^{\mathrm i (\mathbf k_1\cdot \mathbf r-\omega t)}.
\end{align}
Due to the periodicity of the assumed infinite array of ribbons, the EM fields reflected by the grating are given by the Fourier-Floquet decomposition 
\begin{align}
    \mathbf B^\mathrm r (\mathbf r)&=\hat{\mathbf y}\sum _{n=-\infty}^\infty r_n\mathrm e^{\mathrm i (q_nx-\kappa_{1,n}z)}\label{eq:main_B_ref},\\
    \mathbf E^\mathrm r (\mathbf r)&=-\frac{c^2}{\omega\varepsilon_1}\sum _{n=-\infty}^\infty r_n(\hat{\mathbf x}\kappa_{1,n}+\hat{\mathbf z}q_n)\mathrm e^{\mathrm i (q_nx-\kappa_{1,n}z)}, \label{eq:main_E_ref}
\end{align}
with $c$ being the speed of light in vacuum, and the transmitted EM fields propagating in the substrate $\varepsilon_2$ are likewise given by
\begin{align}
    \mathbf B^\mathrm t (\mathbf r)&=\hat{\mathbf y}\sum _{n=-\infty}^\infty t_n\mathrm e^{\mathrm i (q_nx-\kappa_{2,n}z)},\label{eq:main_B_tra}\\
    \mathbf E^\mathrm t (\mathbf r)&=\frac{c^2}{\omega\varepsilon_2}\sum _{n=-\infty}^\infty t_n(\hat{\mathbf x}\kappa_{2,n}-\hat{\mathbf z}q_n)\mathrm e^{\mathrm i (q_nx+\kappa_{2,n}z)}. \label{eq:main_E_tra}
\end{align}
These decompositions are given in terms of Bloch modes of the transverse wave vector component 
\begin{align}
    q_n=k_x+nG,\label{eq:main_Bloch_mode_q}
\end{align}
with $G=2\pi /L$ being the reciprocal lattice constant of the nanoribbons. For the total wave number $\sqrt{\varepsilon_i}k_0$, where $k_0 = \omega/c$ is the free-space wave number, in the medium with permittivity $\varepsilon_i$, the normal wave vector component $\kappa_{i,n}$ in each medium is given by the equation 
\begin{align}
    q_n^2+\kappa_{i,n}^2=\varepsilon_i k_0^2.
\end{align}
The coefficients $r_n$ and $t_n$ associated with the $n$'th Bloch mode are reflection and transmission coefficients, which we will need to obtain the reflectance, transmittance, and absorbance of the nanoribbon. Not all Bloch modes $n$ are propagating modes, in fact most exist only in the near field, and only reflection and transmission coefficients corresponding to propagating Bloch modes contribute to the reflectance, transmittance, and absorbance. At the interface $z=0$, the transmitted electric field component along $x$ produces a current along the nanoribbons, which is described by an optical conductivity $\sigma(x)$ that depends on position $x$ along the grating. Within a unit cell of length $L$, this position-dependent optical conductivity can be written as 
\begin{align}
    \sigma(x)=\sigma(\omega) \Theta (w/2-|x|),\label{eq:main_position_conductivity}
\end{align}
where $\Theta(x)$ is the Heaviside step function and $\sigma(\omega)$ is the optical conductivity of a continuous layer of the 2D material.

\subsection{Edge condition method}
Our starting point is to apply the edge condition method, where the main assumption is that the current perpendicular to the sharp edges of the hBN ribbon at $x=\pm w/2$ is proportional to the distance to the edge, called the edge condition~\cite{Barkeshli2015}. This assumption can be made when we consider small ribbon widths compared to the wavelength, written as $kw<1$. We are able to obtain closed-form expressions for the reflectance, transmittance and absorbance at the cost of information about resonances of higher order than the fundamental dipole-like mode, which are not included in the ansatz. The edge condition gives the current within the 2D semiconductor ribbon at $|x|\leq w/2$ as 
\begin{align}
    \mathbf K(x)=\hat{\mathbf x}\chi \mathrm e^{\mathrm i k_xx}\sqrt{w^2/4-x^2}\Theta(w/2-|x|),\label{eq:main_ansatz}
\end{align}
where $\chi$ is a, for now, unknown coefficient. Application of the boundary conditions with the Fourier-Floquet decomposition of the fields lead to the transmission and reflection coefficients of the nanoribbons as
\begin{align}
    \frac{r_n}{B_0^\mathrm i }&=\delta_{n0}-\frac{\varepsilon_1 \kappa_{2,n}}{\varepsilon_2 \kappa_{1,n}} \frac{t_n}{B_0^\mathrm i },\label{eq:main_r_m}\\
    \frac{t_n}{B_0^\mathrm i } &=\frac{\varepsilon_2\kappa_{1,n}}{ \varepsilon_1\kappa_{2,n}+ \varepsilon_2 \kappa_{1,n}}\left[2\delta_{n0}-\frac{\chi}{B_0^\mathrm i } \frac{\mu_0 w}{4n}J_1\left(\frac{n\pi w}{L}\right)\right],\label{eq:main_t_m}
\end{align}
where $J_1(x)$ is the Bessel function of order 1 and the coefficient $\chi$ is determined as
\begin{align}
    \frac{\chi}{B_0^\mathrm i} =\frac{2\kappa_{2,0}\kappa_{1,0}}{ \varepsilon_1\kappa_{2,0}+ \varepsilon_2 \kappa_{1,0}}\frac{\sigma(\omega)c^2}{\omega } \frac{1}{\Lambda(\omega)},\label{eq:main_chi}
\end{align}
for a frequency-dependent factor with dimensions of length 
\begin{align}
    \Lambda (\omega)= \sum _{n=-\infty }^\infty\frac{w}{4n}J_1\left(\frac{n\pi w}{L}\right)\left[1+\frac{\sigma(\omega)}{\omega \varepsilon_0 } \frac{\kappa_{2,n}\kappa_{1,n}}{ \varepsilon_1\kappa_{2,n}+ \varepsilon_2 \kappa_{1,n}}\right].
\end{align}
This infinite sum in the expression for $\Lambda (\omega)$ is computed for the $n=0$ term and for twice the $n>0$ terms, given that the summand is even with respect to $n$, up to a truncation $n=N$. For this study, a choice of \mbox{$N=101$} leads to well-converged results for the coefficient $\chi$, where the choice of an odd integer $N$ is required for a correct result. The convergence based on the choice of $N$ is further discussed in App.~\ref{sec:conv}. 

For the reflected and transmitted EM waves of a given mode $n$, any energy flux along $z$ is only obtained if the wave number \mbox{$\kappa_{1,n}=\sqrt{\varepsilon_1k_0^2-q_n^2}$} remains real and represents a propagating wave, which is the case for $\varepsilon_1k_0^2>q_n^2$. If, on the other hand, we have $\varepsilon_1k_0^2<q_n^2$, then \mbox{$\kappa_{1,n}=\mathrm i \sqrt{q_n^2- \varepsilon_1k_0^2}$} becomes imaginary and represents an evanescent wave that does not carry any energy along $z$. Similarly, for the transmitted wave, only modes $n$ where \mbox{$\kappa_{2,n}=\sqrt{\varepsilon_2k_0^2-q_n^2}$} remains real and represents a propagating wave contribute to the energy flux. The reflectance, transmittance, and absorbance are given as 
\begin{align}
    R(\omega,\theta)&=\sum_{n\in P}\frac{\mathrm{Re}\{\kappa_{1,n}\}}{\mathrm{Re}\{k_z\}}\left|\frac{r_n}{B_0^\mathrm i }\right|^2,\label{eq:main_R_0}\\
    T(\omega,\theta)&=\sum_{n\in P}\frac{\mathrm{Re}\{\varepsilon_1\kappa_{2,0}\}}{\mathrm{Re}\{\varepsilon_2k_z\}}\left|\frac{t_n}{B_0^\mathrm i }\right|^2,\label{eq:main_T_0}\\
    A(\omega,\theta)&=1-R(\omega,\theta)-T(\omega,\theta),\label{eq:main_A_0}
\end{align}
where the sum in these expressions only includes the propagating modes $n\in P$. 

For the energy window of interest for hBN at \mbox{$\hbar\omega = 4$ eV} to $8$ eV, the condition $kw<1$ limits the ribbon widths to below around 25 nm. At normal incidence $\theta=0$ and for a periodicity given as twice the ribbon width $L=2w$, the relative permittivities in the surrounding dielectrics would need to satisfy the condition
\begin{align}
    \varepsilon_j>\frac{\pi^2c^2}{w^2\omega^2}\label{eq:kw_cond}
\end{align}
in order to obtain propagating modes $n=\pm 1$. As a substrate, we consider a toy encapsulating dielectric of relative permittivity $\varepsilon_1 = 3$ and a toy dielectric substrate of a higher relative permittivity $\varepsilon_2=4$. For these dielectrics, the condition is not satisfied in the energy window of interest for hBN at ribbon widths below 25 nm in neither the substrate nor the surrounding dielectric, meaning that the $n=0$ mode remains the only propagating mode. Similarly, for the energy window of interest $\hbar\omega = 1.5$ eV to $3$ eV for WS\textsubscript{2} nanoribbons, the condition $kw<1$ leads to ribbon widths that are limited to below approximately 66 nm. For this width restriction, a periodicity $L=2w$, and the relative permittivities $\varepsilon_1 = 3$ and $\varepsilon_2=4$, the condition of Eq.~\eqref{eq:kw_cond} is again not satisfied at normal incidence in neither the substrate nor the surrounding dielectric. For a single propagating mode $n=0$, all other modes $n\neq 0$ are imaginary both above and below the 2D semiconductor. 
\begin{figure}[ht]
\centerline{\includegraphics[width=0.9\linewidth]{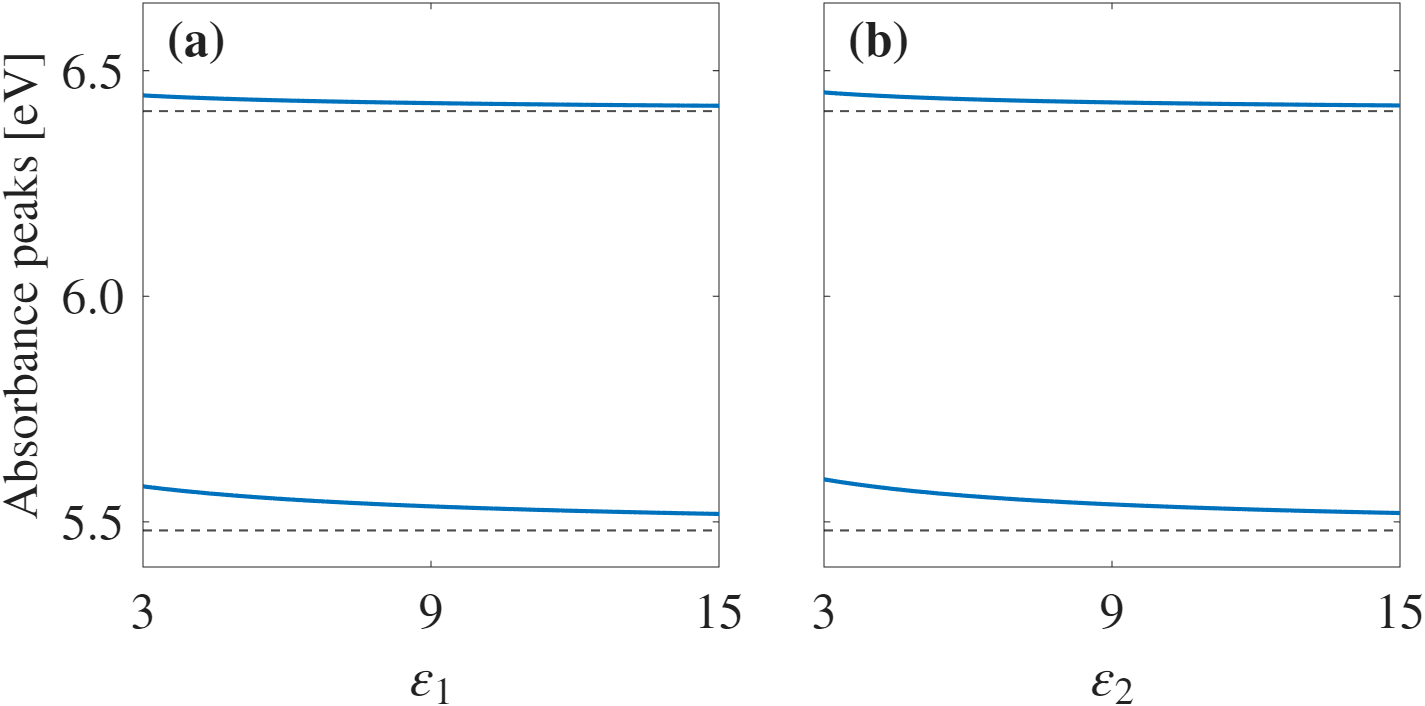}}
\caption{Spectral position of absorbance peaks of hBN nanoribbons calculated using the edge condition method for $w=4\ \mathrm{nm}$ and $w=2L$ for \textbf{(a)} varying $\varepsilon_1$ and fixed $\varepsilon_2=4$ and for \textbf{(b)} varying $\varepsilon_2$ and fixed $\varepsilon_1=3$. The excitonic resonances of hBN are given as the vertical dashed lines.
\label{fig:nanoribbons_vary_eps}}
\end{figure}
The absorbance $A_0$ obtained by Eq.~\eqref{eq:main_A_0} for a single propagating mode $n=0$ of hBN and WS\textsubscript{2} nanoribbons is shown in Fig.~\ref{fig:nanoribbons}\textcolor{blue}{a} and \ref{fig:nanoribbons}\textcolor{blue}{c}, respectively, across a range of ribbon widths $w$ with a periodicity given as $L=2w$. Around the excitonic resonances $\hbar\omega_1$ and $\hbar\omega_2$ for hBN and the $A$ and $B$ lines for WS\textsubscript{2}, increases in absorbance are found across all ribbon widths, but as the ribbon width is decreased, the absorbance peaks slightly blueshift. This shift is caused by the increased confinement of the EM mode, with the shift only manifesting at such small widths demonstrating the localized nature of the excitons. At the atomistically low widths where blueshifts occur, these results should be considered a qualitative feature of the patterning rather than a realistic result of the system, as atomistic modeling of the ribbons would be needed to validate these predictions. Varying the periodicity $L$ of the array at a constant ribbon width of $w=4\ \mathrm{nm}$ for hBN in Fig.~\ref{fig:nanoribbons}\textcolor{blue}{b} and $w=10\ \mathrm{nm}$ for WS\textsubscript{2} in Fig.~\ref{fig:nanoribbons}\textcolor{blue}{d} leads to variations in absorbance while the peaks are still at the excitonic resonances. The WS\textsubscript{2} nanoribbons also show a broader, higher-energy response due to interband transitions. The effect of the surrounding dielectrics is illustrated in Fig.~\ref{fig:nanoribbons_vary_eps}, where the blueshift of the two absorption peaks for hBN nanoribbons is minimized either by increasing the relative permittivity of $\varepsilon_1$ at a fixed $\varepsilon_2=4$ or by increasing $\varepsilon_2$ at a fixed $\varepsilon_1=3$ for $w=4\ \mathrm{nm}$ and $L=2w$, as a result of the increased screening that the excitons in the 2D material experience from their dielectric environment.

\begin{figure}[ht]
\centerline{\includegraphics[width=0.9\linewidth]{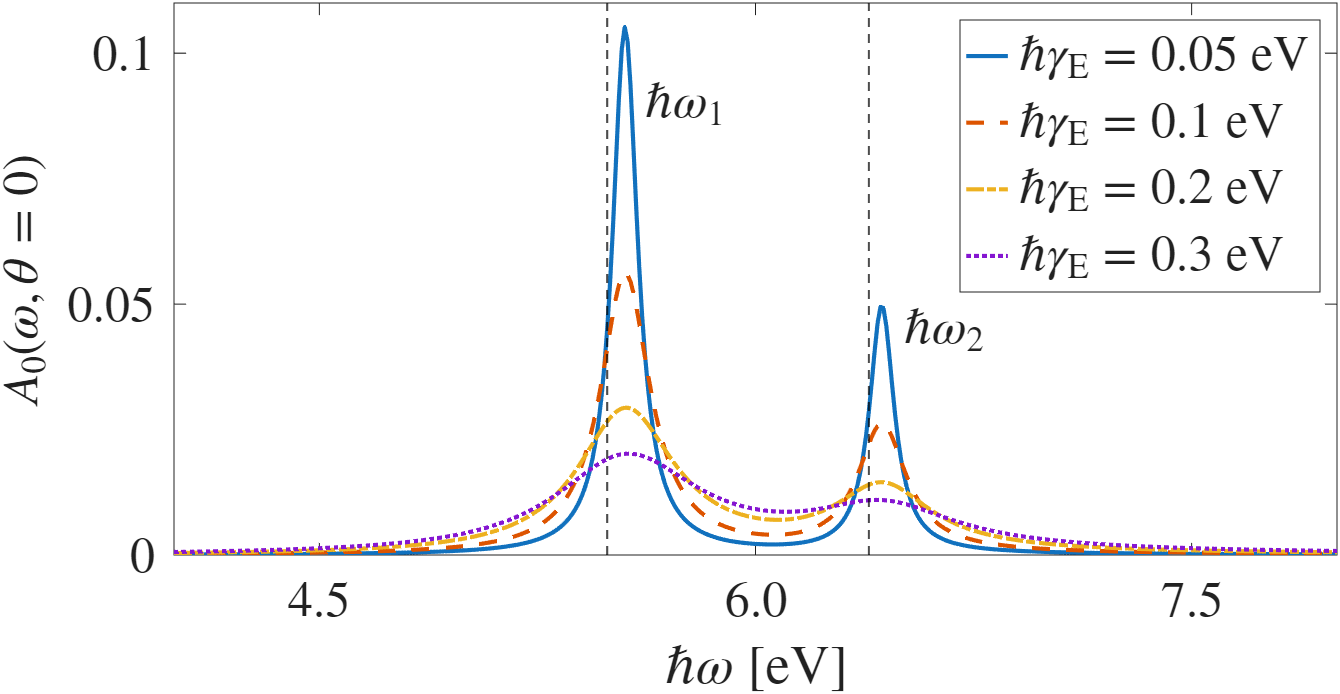}}
\caption{Absorbance of hBN nanoribbons calculated using the edge condition method for $\varepsilon_1=3$, $\varepsilon_2=4$, $w=4\ \mathrm{nm}$ and $L=2w$ at normal incidence for varying linewidth $\gamma_\mathrm{E}$. The excitonic resonances of hBN are given as the vertical dashed lines.
\label{fig:nanoribbons_vary_gE}}
\end{figure}
The impact of the exciton linewidth on the optical response is investigated by using different values of $\gamma_\mathrm E$ in the Elliott equation for hBN. Figure~\ref{fig:nanoribbons_vary_gE} shows the absorption spectrum of hBN nanoribbons for increasing linewidths and fixed $\varepsilon_1=3$, $\varepsilon_2=4$, $w=4\ \mathrm{nm}$ and $L=2w$ at normal incidence, with the only notable effect being a broadening of the resonance peaks and no changes in frequency. This demonstrates that the level of tunability is a result of the geometry while the linewidth only influences resolvability.

Applying the edge condition method results in very poor tunability of the excitonic response by varying the width and periodicity of the ribbons. Our calculations have further shown that this picture does not change with the angle of incidence. To further support this conclusion, we move on to study the same 2D semiconductor nanoribbons using a Fourier series method that does not use an ansatz for the form of the optical conductivity. 

\subsection{Fourier series method}
We now consider a method of writing the optical conductivity of the ribbons as a Fourier series, which requires solving a linear algebra problem at larger computational cost. An important difference from the edge condition method is that, with no ansatz as to the form of the optical conductivity, this method is more complete than the approximate edge condition method, and we are no longer restricted by $kw<1$. Furthermore, this method can be applied to any periodic modulation of the 2D material~\cite{Peres2012_JPhysCondensMatter24}. The decomposition of the optical conductivity is given as
\begin{align}
    \sigma(x)=\sum_{m=-\infty }^\infty \tilde \sigma_m\mathrm e^{\mathrm i mGx},
\end{align}
where the unit cell of the modulated conductivity of Eq.~\eqref{eq:main_position_conductivity} leads to expansion coefficients 
\begin{align}
    \tilde \sigma_0 &= \frac{w}{L}\sigma(\omega)&\mathrm{for}\ m=0,\\
    \tilde \sigma_m &=\frac{1}{\pi m}{\sigma(\omega)}\sin\left(\frac{\pi mw}{L}\right)\mathrm{e}^{\mathrm i \pi mw/L}&\mathrm{for}\ m\neq 0 .
\end{align}
Applying this Fourier decomposition of the optical conductivity and the Fourier-Floquet decomposition of the fields leads to an infinite set of linear equations for all $n$ as
\begin{align}
    \left(\frac{\kappa_{1,n}}{\varepsilon_1 }+\frac{\kappa_{2,n}}{\varepsilon_2 }\right)\frac{t_n}{B_0^\mathrm i}+\frac{\kappa_{1,n}}{\omega \varepsilon_0 \varepsilon_1}\sum_{l=-\infty }^\infty \frac{\kappa_{2,l} }{\varepsilon_2 } \frac{t_l}{B_0^\mathrm i}\tilde \sigma_{n-l}= \frac{2k_z}{\varepsilon_1} \delta_{n0}.\label{eq:main_lineq_n=0}
\end{align}
We now need to truncate both the number of equations $n$ and the number of terms $\ell$ in each equation. For simplicity, both $n$ and $\ell$ are chosen to be truncated as $|n|,|\ell|>N$. For this study, a value of $N=100$ was implemented, with this choice discussed in App.~\ref{sec:conv}. This allows us to set up the set of linear equations as a matrix problem involving a $(2N+1)\times (2N+1)$ matrix containing all the coefficients to the unknown  Bloch components of the transmission coefficients $t_n$. Having solved this matrix problem, the reflection coefficients are given by Eq.~\eqref{eq:main_r_m}, and the reflectance, transmittance and absorbance are again given by Eqs.~\eqref{eq:main_R_0} to \eqref{eq:main_A_0}.

The absorbance $\mathcal A$ of hBN and WS\textsubscript{2} nanoribbons obtained using the Fourier series method when varying the ribbon width is shown in Fig.~\ref{fig:nanoribbons}\textcolor{blue}{e} and \ref{fig:nanoribbons}\textcolor{blue}{g}, respectively, where it is seen to align well with the absorbance obtained using the edge condition method. However, the hBN nanoribbons now exhibit secondary peaks at higher energies as diffraction modes between neighboring ribbons. Figure~\ref{fig:nanoribbons}\textcolor{blue}{f} shows that as the periodicity is decreased at a fixed width, i.e. as the ribbons are brought closer together, these higher-energy peaks are attenuated. These peaks are not seen for the WS\textsubscript{2} nanoribbons, as the oscillator strengths of the excitons are lower than for hBN, and the higher-energy diffraction orders may be suppressed by the interband response. As higher-order diffraction modes are not implemented into the ansatz of the current density in the edge condition method, this small peak is not present in Fig.~\ref{fig:nanoribbons}\textcolor{blue}{a} this method does not show such diffraction orders. The higher oscillator strength of hBN is manifested as a larger absorption in the nanoribbons as compared to the WS\textsubscript{2} nanoribbons. Unlike in graphene, where collective plasmons dominate the response, and these are very much geometry dependent, here we see how surface exciton polariton effects show little tunability with the dimensions of the ribbons due to their localized nature, meaning that the coupling and hybridization of the excitons with an optical mode give very little freedom. This geometry-independence of the excitonic response of 2D semiconductor ribbons---as compared to graphene nanoribbons---is interesting in its own right, and has, to the best of our knowledge, not been studied in literature, although efforts related to a periodic external modulation have been gaining interest in the last couple of years~\cite{Ninhos2024_ACSNano18, Lassaline2025_NanoLett25, Danielsen2025_ACSNano19}.

\begin{figure}[ht]
\centerline{\includegraphics[width=0.9\linewidth]{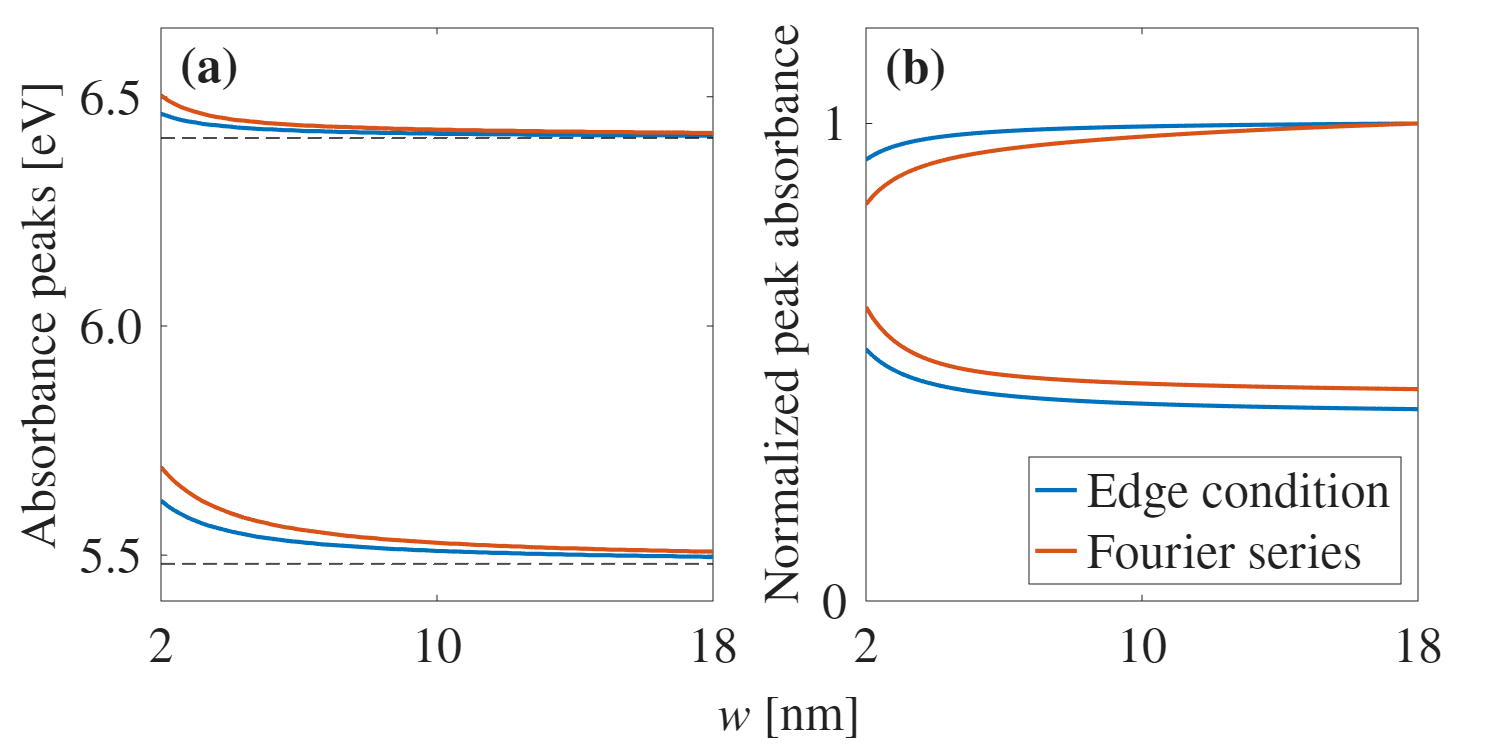}}
\caption{Comparison of the edge condition and Fourier series method when applied to hBN nanoribbons of varying width $w$ with a periodicity of $L=2w$ in an environment $\varepsilon_1=3$ and substrate
$\varepsilon_2=4$ for radiation of normal incidence $\theta=0$. \textbf{(a)} Spectral position of the two main peaks and \textbf{(b)} normalized peak values in absorbance for the edge condition (Fourier series) method as the blue (red) curves. \label{fig:nanoribbons_compare}}
\end{figure}
The two main absorbance peaks differ slightly in their spectral positions between the two methods at lower ribbon widths. For hBN nanoribbons, the spectral positions of the two peaks are plotted in Fig.~\ref{fig:nanoribbons_compare}\textcolor{blue}{a} when using the edge condition (Fourier series) method as the two blue (red) curves. While both methods capture the blueshift, the Fourier series method predicts a less significant shift. The Fourier series method predicts higher absorbance values for the two main peaks, as seen in Fig.~\ref{fig:nanoribbons_compare}\textcolor{blue}{b}. The upper curve for each method corresponds to the lower-energy peak around the excitonic resonance $\hbar\omega_1$, as this peak shows a higher absorbance than the higher-energy peak around the excitonic resonance $\hbar\omega_2$. The blue (red) curves for the edge condition (Fourier series) method are normalized to the maximum absorbance obtained for the lower-energy peak at a ribbon width of $w=18\ \mathrm{nm}$. The Fourier series method illustrates how the edge condition method tends to underestimate the tunability of the response for 2D semiconductor ribbons. However, especially at the larger ribbon widths investigated in this work, the differences in blueshift between the two methods and the contribution of higher-order diffraction peaks become negligible, which justifies the use of the approximate edge condition method at a lower computational cost.

To further pursue exclusively geometry-based efficient patterning, we now move on to describe spherical systems which have the potential to enable tunability of the excitonic response by either excitation of surface-exciton polaritons, or via coupling of the excitons to the modes of a dielectric particle.
\begin{figure}[h]
\centerline{\includegraphics[width=0.9\linewidth]{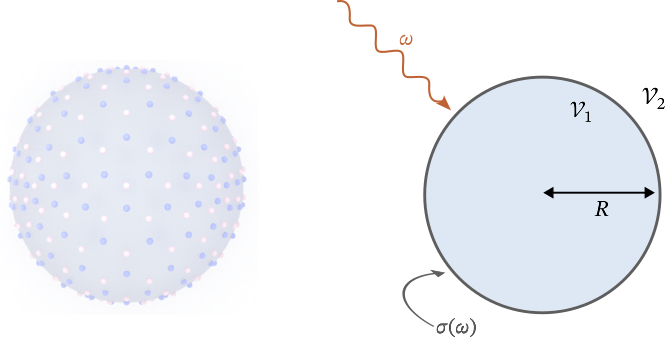}}
\caption{Illustration (left) and schematic (right) of the system described in Sec. \ref{sec3} consisting of a spherical particle of radius $R$ and a relative permittivity $\varepsilon_1$ coated by a 2D material of conductivity $\sigma(\omega)$. The particle is placed in a medium of relative permittivity $\varepsilon_2$ and is illuminated by a plane wave of angular frequency $\omega$. \label{fig:particle}}
\end{figure} 

\section{Localized surface excitons in spherical NPs}\label{sec3}
We now consider the optical response of a spherical NP coated with a 2D semiconductor and illuminated by a monochromatic plane wave, as illustrated in Fig.~\ref{fig:particle}. The coupling of the external radiation with the excitons of the 2D semiconductor leads to localized exciton polaritons~\cite{Gentile2016_JOpt18, Gentile2017_JOpt19,Dutta2024_PhysRevB109}. The formalism we apply to describe the response of spheres at the nanoscale was developed in 1908 by Gustav Mie~\cite{Mie1908_AnnPhys330}. Derivations for conventional particles in the absence of 2D materials can be found in textbooks~\cite{Hohenester2019, Bohren2008}, and Mie theory, modified to incorporate a 2D material as a surface conductivity, has been investigated earlier for the case of graphene~\cite{Christensen2015_PhysRevB91}. As the NP is illuminated, the charge carriers of the material, which are either free or bound depending on the nature of the material, are driven in oscillation by the electric field. The acceleration of these electric charges leads to EM energy radiating away from the particle, a mechanism known as scattering. However, through damping mechanisms such as collisions of the excitons, the motion of the charge carriers also leads to the transformation of the incident energy into other forms of non-radiative energy; a mechanism known as absorption. What follows is a description of Mie theory modified to include 2D materials, still assumed to be infinitely thin as to not affect the radius of the NP, at the surface of the particle. 

\subsection{Modified Mie theory applied to 2D materials}
The particle of radius $R$ occupies a spherical region $\mathcal V_1$ and is placed in a homogeneous host region $\mathcal V_2$, taken as a dielectric material of relative permittivity $\varepsilon_2$. Wave numbers in these regions are denoted as $k_i=\sqrt{\varepsilon_i}k_0$ for $i=1,2$. We assume that the relative permeability of each material is given as $\mu_i=1$ for $i=1,2$ in the optical regime. The analysis is limited to linear materials, and we work within the local response approximation. The incident EM fields $\mathbf E_{\mathrm{inc}},\mathbf H_{\mathrm{inc}}$, along with the fields inside the particle $\mathbf E_{1},\mathbf H_{1}$ and the fields scattered by the particle $\mathbf E_{\mathrm{sc}},\mathbf H_{\mathrm{sc}}$, are given as multipole expansions in the form
\begin{widetext}
\begin{align}
    \mathbf E(\mathbf r)&=\sum_{\ell=1}^\infty \sum_{m=-\ell}^\ell \left[\frac{\mathrm i }{k_i}a_{E\ell m}\boldsymbol \nabla \times f_\ell(k_i r)\mathbf X_{\ell m }(\hat{\mathbf r})+a_{H\ell m}f_\ell(k_i r)\mathbf X_{\ell m }(\hat{\mathbf r})\right],\\
    \mathbf H(\mathbf r)&=\sqrt{\frac{\varepsilon_i\varepsilon_0}{\mu_i\mu_0}}\sum_{\ell=1}^\infty \sum_{m=-\ell}^\ell \left[ a_{E\ell m}f_\ell (k_i r)\mathbf X_{\ell m }(\hat{\mathbf r})-\frac{\mathrm i}{k_i}a_{H\ell m}\boldsymbol \nabla \times f_\ell (k_i r)\mathbf X_{\ell m }(\hat{\mathbf r})\right]
    \normalsize
\end{align}
\end{widetext}
for unknown coefficients $a_{P\ell m}$ where $P=E,H$ is the polarization, and for functions of arguments $f = f(k_i r)$ for $i=1,2$. These expansions are given in terms of vector spherical harmonics $\mathbf X_{\ell m}(\hat{\mathbf r})$ of order $\ell$ and $m$~\cite{Jackson1998}, where $\hat{\mathbf{r}}$ denotes the solid angle. Each term for a given value of $\ell$ corresponds to the EM fields of a multipole, with $\ell=1$ being dipole contributions to the total fields, $\ell = 2$ being quadrupole contributions, and so on. For the expansions of the incident EM fields, the unknown coefficients are denoted as $a^0_{P\ell m}$, and the functions $f_\ell (x)$ are the spherical Bessel functions $j_\ell (x)$ of order $\ell$. For the expansions of the EM fields inside the particle, the functions $f_\ell (x)$ are again the spherical Bessel functions. The asymptotic form of spherical Bessel functions of large arguments is appropriate to describe the localized exciton polaritons as standing spherical waves inside the particle~\cite{Arfken2011}. For the expansions of the scattered EM fields, the unknown coefficients are denoted as $a^+_{P\ell m}$. In the region $\mathcal V_2$, the scattered-field expansions can be written in terms of spherical Hankel functions of the first kind $h_\ell (x)$. The asymptotic form of spherical Hankel functions of large arguments are appropriate to describe propagating spherical waves outside the particle. 

\begin{figure*}[ht]
\centerline{\includegraphics[width=\linewidth]{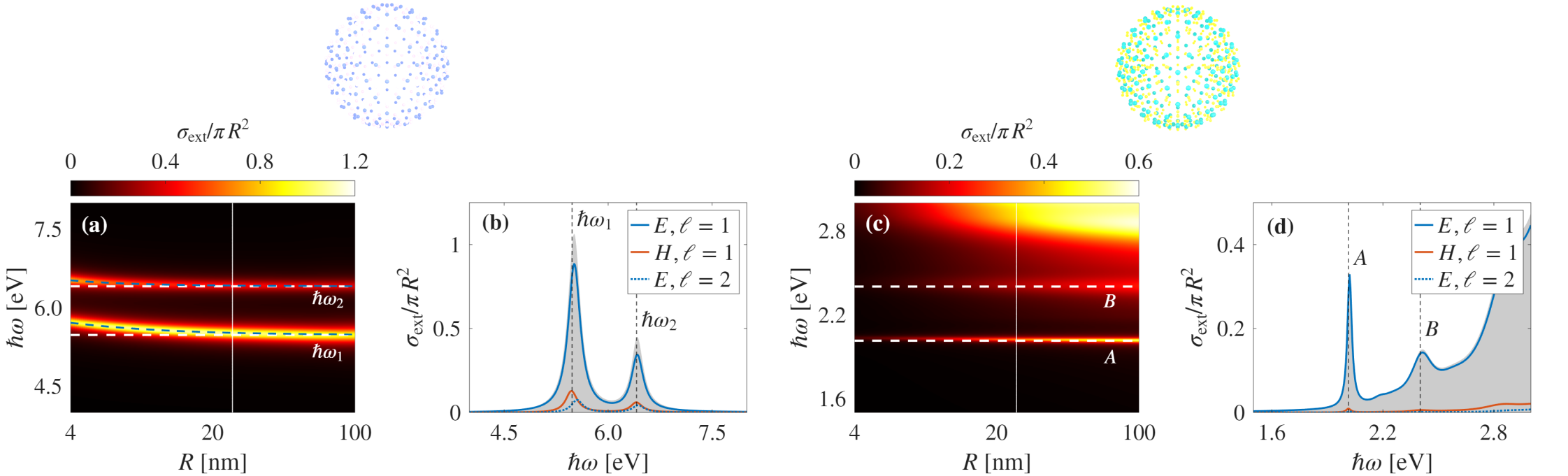}}
\caption{Extinction cross-section, normalized to the geometric cross-section, of a hollow nanoshell of 2D semiconductors surrounded by air $\varepsilon_1=\varepsilon_2=1$. \textbf{(a)} Extinction cross-section of a hBN shell as a function of the radius $R$. The excitonic resonances $\hbar\omega_1$ and $\hbar\omega_2$ of monolayer hBN are shown as the horizontal white dashed lines, and the quasistatic electric dipole resonances of Eq.~\eqref{eq:Mie_approx_res_pm} are shown as the blue dashed curves. \textbf{(b)} Extinction cross-section of a hBN shell at $R=25\ \mathrm{nm}$. The electric dipole (quadrupole) contribution is shown as the solid (dashed) blue line, and the magnetic dipole contribution is shown as the solid red line, with the extinction cross-section given as the shaded gray area. \textbf{(c)} Extinction cross-section of a WS\textsubscript{2} shell as a function of the radius $R$. The $A$ and $B$ peaks are shown as the horizontal white dashed lines. \textbf{(d)} Extinction cross-section of a WS\textsubscript{2} shell at $R=25\ \mathrm{nm}$.
\label{fig:nanospheres}}
\end{figure*}

The total EM field outside the particle consists of both the incident fields and the scattered fields as \mbox{$\mathbf E_{2}=\mathbf E_{\mathrm{inc}}+\mathbf E_{\mathrm{sc}}$} and \mbox{$\mathbf H_{2}=\mathbf H_{\mathrm{inc}}+\mathbf H_{\mathrm{sc}}$}. With a discontinuity of the tangential magnetic fields across the interface from the introduction of the 2D coating, the scattering Mie coefficients, being the ratios of unknown coefficients \mbox{$T_{P\ell }=a_{P\ell m }^+/a_{P\ell m}^0$} for $P=E,H$, are given as
\begin{widetext}
\begin{align}
    T_{E\ell }&=\frac{j_{\ell}(x_1)\Psi'_{\ell}(x_2)\varepsilon_1-\Psi'_{\ell}(x_1)j_{\ell}(x_2)\varepsilon_2+g (\omega)\Psi '_\ell (x_1)\Psi '_\ell (x_2)}{\Psi'_{\ell}(x_1)h_{\ell}(x_2)\varepsilon_2-j_{\ell}(x_1)\xi'_{\ell}(x_2)\varepsilon_{1}-g (\omega)\Psi '_\ell (x_1)\xi '_\ell (x_2)}\label{TEell},\\
    T_{H\ell }&=\frac{j_{\ell}(x_1)\Psi'_{\ell}(x_2)-\Psi'_{\ell}(x_1)j_{\ell}(x_2)+g (\omega)x_0^2j_\ell (x_1)j_\ell (x_2)}{\Psi'_{\ell}(x_1)h_{\ell}(x_2)-j_{\ell}(x_1)\xi'_{\ell}(x_2)-g (\omega)x_0^2j_\ell (x_1)h_\ell (x_2)}.\label{THell}
\end{align}
\end{widetext}
Here, we have introduced dimensionless variables \mbox{$x_i=k_i R$} for $i=0,1,2$, along with $\Psi_\ell '(x)$ and $\xi_\ell '(x)$ as the derivatives of the Ricatti-Bessel functions $\Psi_\ell (x)=xj_\ell (x)$ and $\xi_\ell (x)=xh_\ell (x)$ with respect to their argument $x$. The surface conductivity due to the 2D coating is given by the dimensionless parameter 
\begin{align}
    g(\omega) = \frac{\mathrm i \sigma (\omega)}{\varepsilon_0 \omega R}, \label{eq:g(w)}
\end{align}
entering the third term of the numerator and denominator of both Mie coefficients. The conventional Mie coefficients are therefore restored by setting $g(\omega) = 0$. This is analogous to methods for including nonlocality in the NP description, where a surface response function, rather than a surface conductivity, appears in additional terms to both the numerator and the denominator of the conventional Mie coefficients~\cite{Bundgaard2024_JOptSocAmB41, Kyvelos2025_JPhysChemC129}. 

For NPs that are small compared to the wavelength of the incident radiation, at a given time any point within the particle experiences approximately the same phase of the incident EM fields, which is known as the quasistatic regime. This is compared to larger particles, where two points within the particle might be so separated as to experience the same field with a retardation between the two points. The quasistatic limit is given for radial points $r$ such that $k_ir\ll 1$ both inside and outside the particle, $i=1,2$, allowing for the spherical Bessel functions and spherical Hankel functions of the first kind to be written in their asymptotic forms~\cite{Arfken2011}, leading to the approximate Mie coefficients
\begin{align}
    T_{E\ell } &\simeq  \frac{\mathrm i(\ell +1)x_2^{2\ell +1}}{(2\ell +1)!!(2\ell -1)!!}\frac{\varepsilon _1-\varepsilon _2+(\ell +1)g (\omega)}{(\ell +1)\varepsilon _2+\ell \varepsilon _1+\ell (\ell +1)g(\omega)},\label{eq:quasistatic_electric_Mie}\\
    T_{H\ell } &\simeq  \frac{\mathrm i(\ell +1)x_2^{2\ell +1}}{(2\ell +1)!!(2\ell -1)!!}\frac{(\ell +1)^{-1}x_0^2g (\omega)}{2\ell +1 -x_0^2g(\omega)}.
\end{align}

The scattering cross-section of the particle is the ratio of the power scattered by the particle relative to the incident energy flow normal to the direction of propagation~\cite{Hohenester2019}. Normalized to the geometric cross-section $\pi R^2$ of the particle it is given in terms of the scattering Mie coefficients as 
\begin{align}
    \sigma _\mathrm{sc}=&\frac{2 }{(k_2R)^2}\sum_{\ell=1}^\infty (2\ell +1)(|T_{E\ell }|^2+|T_{H\ell }|^2).\label{eq:sigma_sc}
\end{align}
The extinction is interpreted in experiments as the energy lost from the incident radiation when measuring the resulting radiation on the opposite side of the particle. Similarly, the extinction cross-section, normalized to the geometric cross-section, is given in terms of the scattering Mie coefficients as
\begin{align}
    \sigma_\mathrm{ext}=-\frac{2 }{(k_2R)^2}\sum_{\ell =1}^\infty (2\ell +1)\mathrm{Re}(T_{E\ell }+T_{H\ell }). \label{eq:sigma_ext}
\end{align}
The normalized absorption cross-section can then be determined through the equation $\sigma_\mathrm{abs} = \sigma_\mathrm{ext} - \sigma_\mathrm{sc}$.

\begin{figure*}[ht]
\centerline{\includegraphics[width=\linewidth]{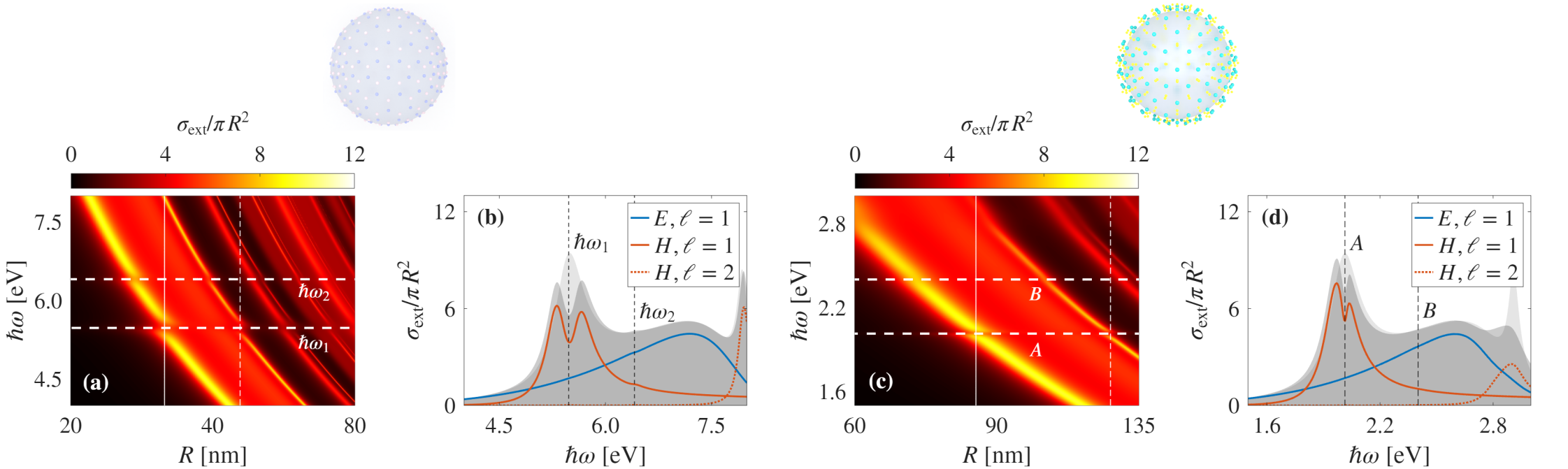}}
\caption{Extinction cross-section, normalized to the geometric cross-section, of a dielectric core $\varepsilon_1=11.7$ coated by a 2D semiconductor surrounded by air $\varepsilon_2=1$. \textbf{(a)} Extinction cross-section of a hBN coating as a function of the radius $R$. The excitonic resonances $\hbar\omega_1$ and $\hbar\omega_2$ are shown as the horizontal white dashed lines. \textbf{(b)} Extinction cross-section of a hBN coating at $R=31.6\ \mathrm{nm}$. The electric dipole contribution is shown as the solid blue line, and the magnetic dipole (quadrupole) contribution is shown as the solid (dashed) red line, with the extinction cross-section of the bare dielectric core given as the shaded light-gray area and the extinction cross-section of the hBN coated particle given as the shaded dark-gray area. \textbf{(c)} Extinction cross-section of a WS\textsubscript{2} coating as a function of the radius $R$. The $A$ and $B$ peaks are shown as the horizontal white dashed lines. \textbf{(d)} Extinction cross-section of a WS\textsubscript{2} coating at $R=86.2\ \mathrm{nm}$.}
\label{fig:nanospheres_Si}
\end{figure*}

\subsection{Hollow shells of hBN and WS\textsubscript{2}}
Our starting point is a theoretical investigation of hollow shells, where the 2D semiconductor is placed in vacuum \mbox{$\varepsilon_1=\varepsilon_2=1$}. This is the simplest possible example of implementing 2D materials into a spherical geometry, and is considered in order to highlight the excitonic response of 2D semiconductors. The spherical geometry also has experimental relevance, with carbon atoms being capable of forming buckminsterfullerenes~\cite{Kroto1985_Nature318} along with being a NP coating~\cite{Yang2014_CeramicsInternational40,Lee2013_ACSNano7}. Such fullerene structure is not limited to carbon, with boron fullerenes~\cite{Zhai2014_NatureChem6} and hBN NPs~\cite{Salles2012_JEurCeramSoc32,Tang2008_AdvFunctMater18} being experimentally feasible. We first consider the extinction cross-section of hollow hBN and WS\textsubscript{2} shells in Fig.~\ref{fig:nanospheres}\textcolor{blue}{a} and \ref{fig:nanospheres}\textcolor{blue}{c}, respectively, across a range of radii and energies. For each radius, we see two clear resonances, which are localized exciton polaritons~\cite{gentile_jopt19}, near the excitonic resonances $\hbar\omega_1$ and $\hbar\omega_2$ for hBN and the $A$ and $B$ lines for WS\textsubscript{2}. The hollow WS\textsubscript{2} shell also shows a higher-energy resonance due to the external field interacting with the interband transitions. By considering the individual terms of $T_{E\ell }$ and $T_{H\ell }$ for any $\ell$ in Eq.~\eqref{eq:sigma_ext}, we are able to identify each electric or magnetic multipole contribution to the total extinction cross-section individually. As illustrated in Fig.~\ref{fig:nanospheres}\textcolor{blue}{b} for hBN and Fig.~\ref{fig:nanospheres}\textcolor{blue}{d} for WS\textsubscript{2} at a radius of $R=25\ \mathrm{nm}$, the extinction cross-section, shown as the shaded gray area, mainly consists of an electric dipole contribution, given by the solid blue curve. Near the excitonic resonances, the driving fields cause collective oscillations of the excitons in the sphere with little to no retardation in order to form this electric dipole contribution. We also see a much less dominant magnetic dipole contribution, which appears due to the current introduced with the hBN layer at the surface of the particle. As the NP size is increased, the size of the current loop also increases, inducing a larger magnetic dipole. Due to retardation effects along the shell, we also see a small electric quadrupole contribution. 

From a theoretical point of view, it is interesting to consider radii of the hollow hBN shell down to $R=4\ \mathrm{nm}$, where both resonances start to shift to higher energies with a decreasing radius. In the limit of such small particles, an atomistic model of the particles would lead to more realistic results, but is beyond the scope of this article. The spectral positions of the extinction peaks for the hBN shell at low radii are predicted from the poles of the scattering Mie coefficients in the quasistatic limit. Given that the electric dipole is the dominating contribution to the extinction, we focus our attention on the poles of the electric scattering Mie coefficient for $\ell = 1$. For the hBN shell described by Eq.~\eqref{eq:cond:hBN} in any dielectric environment, given by $\varepsilon_1$ and $\varepsilon_2$, not restricted to vacuum, we obtain two electric dipole resonances due to the two terms of the Elliot equation. These two electric dipole resonances are approximately given as
\begin{align}
    &\omega_\pm \simeq \frac{1}{2}(\omega_1+\omega_2)+\frac{\omega_\mathrm E(p_1+p_2)}{{(2\varepsilon _2+\varepsilon _1)}}\notag\\
    &\pm \left[\frac{1}{2}\right.\left.|\omega_1-\omega_2|-\frac{\omega_\mathrm{E}(p_1-p_2)}{{(2\varepsilon _2+\varepsilon _1)}}+\frac{\omega_\mathrm{E}^2(p_1+p_2)^2}{({2\varepsilon _2+\varepsilon _1})^2|\omega_1-\omega_2|}\right].\label{eq:Mie_approx_res_pm}
\end{align}
This equation positions the electric dipole resonances near the monolayer hBN excitonic resonances as $\omega_-=\omega_1$ and \mbox{$\omega_+=\omega_2$} but predicts a shift towards higher energies for a decreasing radius for both electric dipole resonances through the last two terms in the bracket. This shift can be explained by an increased screening, which is caused by the introduced surface current in the 2D material, as the radius is decreased. This approximation is shown in Fig.~\ref{fig:nanospheres}\textcolor{blue}{a} as the dashed blue lines, showing excellent agreement with the full solution.

Unlike in hBN, we do not see a pronounced blueshift of the two extinction peaks for the hollow WS\textsubscript{2} shell. Given the higher background permittivity of WS\textsubscript{2} at $\varepsilon_\infty=10.68$, as the radius of the NP decreases, the resonance frequency does not need to increase much for the function $g(\omega)$ of Eq.~\eqref{eq:g(w)} to remain at a value that leads to poles in Eq.~\eqref{eq:quasistatic_electric_Mie} for $\ell = 1$. So far, we see an excitonic response of the spherical 2D semiconducting coating comparable to that of 2D nanoribbons, albeit with a more pronounced possibility for tuning with the radius of the particle compared to the width of the nanoribbons. We will now deposit the semiconducting coating on a dielectric particle in order to investigate the coupling of the excitons to its Mie optical modes.

\subsection{Dielectric particles with conductive coating}

As a means to a more realistic fabrication scenario, instead of a hollow shell, we now consider a toy dielectric core of $\varepsilon_1=11.7$ with a coating of the 2D material, placed in vacuum ($\varepsilon_2=1$). A dielectric core has no free charge carriers that respond to the external radiation. Instead, the external field causes polarization of the molecules that make up the dielectric, leading to local pairs of polarization charges. For large enough dielectric cores, retardation effects cause these polarization charges to oscillate out of phase, which induces a displacement current. Hence it is possible for a dielectric core to support both electric and magnetic multipoles. By introducing a conductive layer to the surface of the dielectric core, we allow for interaction between the excitons in the 2D material and the polarization charges, causing a coupling of their oscillations and the formation of Mie-exciton polaritons~\cite{Todisco2020_Nanophotonics9, Wang2020_JPhysChemC124}. 

The extinction cross-section for the dielectric particle with a hBN coating and a WS\textsubscript{2} coating as a function of radius is seen in Fig.~\ref{fig:nanospheres_Si}\textcolor{blue}{a} and Fig.~\ref{fig:nanospheres_Si}\textcolor{blue}{c}, respectively, where the excitonic resonances are indicated by the dashed white lines. Turning to a radius of \mbox{$R=31.6\ \mathrm{nm}$} for the hBN coated particle in Fig.~\ref{fig:nanospheres_Si}\textcolor{blue}{b} and $R=86.2\ \mathrm{nm}$ for the WS\textsubscript{2} coated particle in Fig.~\ref{fig:nanospheres_Si}\textcolor{blue}{d}, we show the extinction cross-section of the dielectric core with a coating as the dark-gray shaded area, compared to the extinction cross-section of just the bare dielectric core as the light-gray shaded area. At these radii, a magnetic dipole peak for the bare dielectric core aligns with the lower excitonic resonance $\hbar\omega_1$ of hBN and the $A$ line of WS\textsubscript{2}, which causes a splitting of the extinction resonance for the coated particle into two hybrid modes~\cite{tserkezis_rpp83} around the exciton resonance. 
The possibility of such systems entering the strong-coupling regime has been discussed in recent literature~\cite{Tserkezis2018_PhysRevB98,Lepeshov2018_ACSApplMaterInterf10,dutta_prb109}, and the parameters used here are quite similar. A common criterion for strong coupling states that the energy split must be larger than the square root of half the sum of the uncoupled linewidths squared~\cite{Torma2015_RepProgPhys78}. At $R=31.6\ \mathrm{nm}$ for the hBN coated particle, the split of 0.35 eV between the peaks of the hybrid modes does not satisfy this criterion due to the large width of the magnetic dipole contribution from the core. However, at a radius $R=45.7\ \mathrm{nm}$ shown in Fig.~\ref{fig:nanospheres_Si}\textcolor{blue}{a} as the vertical dashed white line, the splitting around the $\hbar\omega_1$ peak takes a value of 0.25 eV, and the uncoupled linewidths of the extinction cross section peaks for the bare hBN sphere and the magnetic quadrupole contribution to the bare dielectric core are found to be 0.24 eV and 0.13 eV, respectively. The linewidths are obtained using a least-squares fitting of the bare peaks as Lorentzians at a window of $\pm 0.2$ eV around $\hbar\omega_1$, where the bare peaks can be approximated as Lorentzians. These values satisfy the condition for strong coupling as \mbox{$\sqrt{(0.24\ \mathrm{eV})^2 + (0.13\ \mathrm{eV})^2}/\sqrt{2} = 0.19\ \mathrm{eV} <0.25\ \mathrm{eV}$}. This demonstrates how the tunability of the nanoparticle is an important factor in the interaction between the excitons with the core. Similarly for the WS\textsubscript{2} coated particle, the split of 0.068 eV due to the interaction of the $A$ peak with the magnetic dipole of the core does not satisfy the criterion for strong coupling, while at $R=124.5\ \mathrm{nm}$, indicated in Fig.~\ref{fig:nanospheres_Si}\textcolor{blue}{c} as the vertical dashed white line, the split of 0.046 eV and uncoupled linewidths of 0.030 eV for the bare WS\textsubscript{2} peak and 0.047 eV for the magnetic quadrupole contribution (obtained by a least-squares fitting in a window of $\pm 0.1$ eV around the $A$ line) does show strong coupling.

At higher energies of around 2.8 eV, we also see that a peak in the bare dielectric core extinction cross-section due to the magnetic quadrupole contribution is suppressed once the WS\textsubscript{2} coating is introduced. This is likely due to interactions of the dielectric with the excitons due to higher-lying interband transitions in WS\textsubscript{2}, which occur at these energies. 

Unlike in the 2D semiconductor nanoribbons, when turning to spherical particles, we both observe much more pronounced tunability with the geometry and now also a hybridization with an introduced dielectric core, leading to a splitting of the Mie resonances at specific particle radii~\cite{Tserkezis2018_PhysRevB98,Lepeshov2018_ACSApplMaterInterf10}.
Similar possibilities open, of course, also in the context of
so-called plexcitonics, with localized surface plasmons in
metals playing the role of the Mie modes~\cite{Fofang2008_NanoLett8}.

\section{Conclusions}
\label{sec5}
We have applied EM methods for determining the excitonic behavior of patterned 2D semiconductors, focusing on hBN in the UV regime and on WS\textsubscript{2} in the visible, patterned into an array of ribbons or surrounding spherical particles. When considering periodic ribbons of a 2D material, absorption peaks are observed for hBN ribbons near the intrinsic excitonic resonances of hBN, which are seen to shift to slightly higher energies for lower ribbon widths. Furthermore, by applying the Fourier series method, a higher diffraction order appears as a higher-energy absorption peak in the spectrum of hBN ribbons. Similarly, for WS\textsubscript{2}, absorption peaks appear very near the excitonic resonances ($A$ and $B$ peaks), exhibiting small shifts in energy for varying ribbon widths. For a spherical geometry, both hBN shells and WS\textsubscript{2} shells show extinction peaks very near the excitonic resonances of a planar sheet of hBN and the $A$ and $B$ peaks of WS\textsubscript{2}, respectively. When coating a dielectric or metallic core with a 2D material, we observe interactions of the excitons of the 2D material with the core, leading to splitting in the extinction resonances with the bare core. The spherical geometry therefore indicates a more fruitful path towards tuning and engineering the hybrid optical response of 2D semiconductors, enabling systematic control through geometric and material parameters. Such tunability opens new paths for tailoring light--matter interactions at the extreme nanoscale, thus enabling the design of miniaturized polaritonic devices. This work should provide a qualitative overview of the expected excitonic features of such devices, even for more realistic dielectric environments than the toy dielectrics investigated. The paper demonstrates the ease and versatility of applying the discussed methods, only requiring an optical conductivity for the 2D material in question to enable investigations of the optical response of the designed nanostructure. It should be noted that while this paper focuses on excitons dominating the optical conductivities of materials, any mechanism leading to resonances in the optical conductivity of a given material can be readily studied using the given methods, leading to the emergence of the corresponding polariton (e.g. exciton- or phonon-polaritons). This allows for a broad class of hybrid light–matter phenomena to be explored within the same framework.

\section*{AUTHOR CONTRIBUTIONS}
CT and LJ conceived the idea and supervised the project. CNH performed analytic calculations,
prepared the computational codes, obtained and plotted the results.
All authors discussed the results and contributed to the writing 
of the manuscript.

\begin{acknowledgments}
We thank Joel D. Cox and Nuno M. R. Peres for discussions and proofreading the manuscript. 

\section*{FINANCIAL DISCLOSURE}
The Center for Polariton-driven Light–Matter
Interactions (POLIMA) is funded by the Danish National
Research Foundation (Project No. DNRF165).

\section*{CONFLICTS OF INTEREST}
The authors declare no conflicts of interest.

%

\end{acknowledgments}

\appendix
\section{Convergence of edge condition and Fourier series methods}
\label{sec:conv}

\begin{figure}[ht]
\centerline{\includegraphics[width=0.9\linewidth]{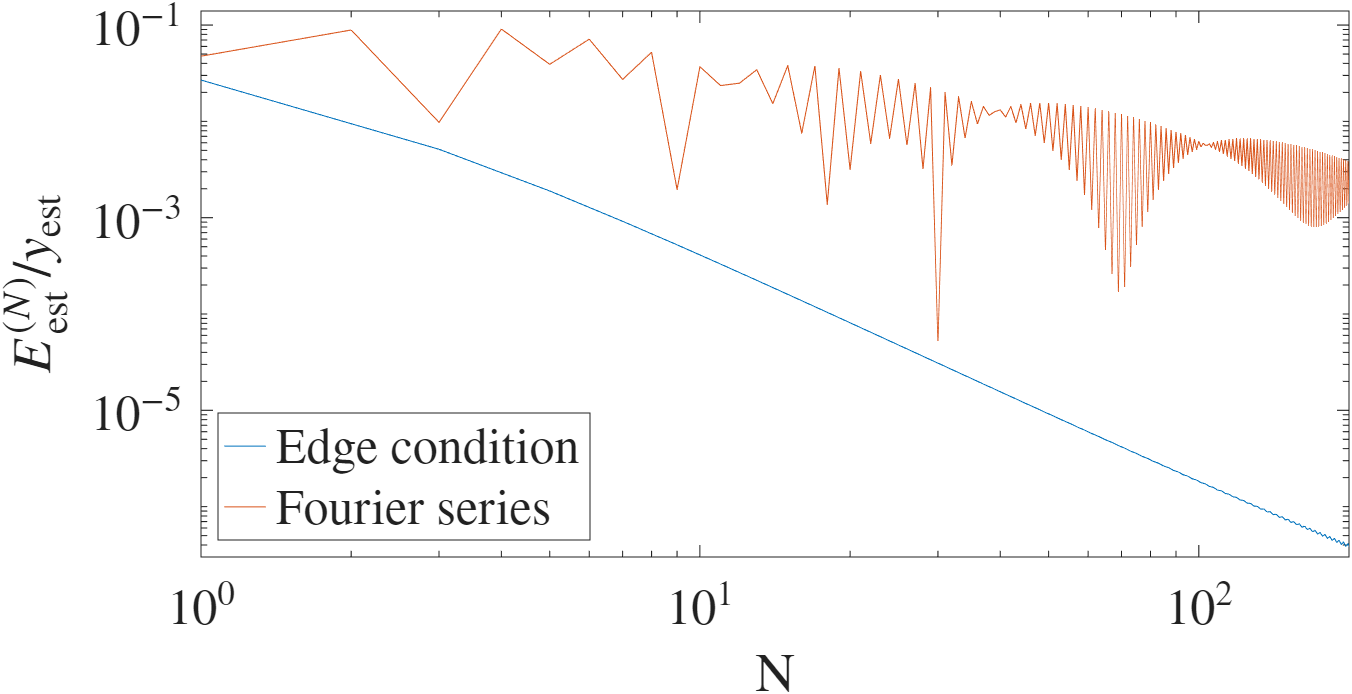}}
\caption{Estimated error $E_\mathrm{est}^{(N)}$, normalized to the true peak value $y_\mathrm{est}$ as a function of $N$ for the edge condition (Fourier series) method shown as the blue (red) curve.}
\label{fig:convergence}
\end{figure}

For both periodic ribbon methods, we consider the hBN nanoribbons of $\varepsilon_1=3$, $\varepsilon_2=4$, $w=4\ \mathrm{nm}$ and $L=2w$ at normal incidence, and consider the value of the absorbance peak of the lower exciton resonance $\hbar\omega_1$ as the figure of merit for evaluating the accuracy of the respective methods. Denoting this peak value as $\tilde y^{(N)}$ for a given $N$, we calculate the estimate of the error as 
\begin{align}
    E_\mathrm{est}^{(N)} = |\tilde y^{(N)} - y_\mathrm{est}|,
\end{align}
For both methods, it is assumed that the obtained peak value $\tilde y^{(1001)}$ is an acceptable estimate of the true peak value $y_\mathrm{est}$ of the given method. For the edge condition method, following the discussion in Ref.~\cite{Goncalves2016_PhysRevB94}, only odd values of $N$ are considered for the convergence analysis. The estimated errors, normalized to the true peak value, of both methods are shown in Fig.~\ref{fig:convergence} for increasing integer value of $N$. We see that the edge condition method converges to its estimated true peak value at much lower $N$, in addition to being much less computationally expensive, than the Fourier series method. The choice of $N=101$ for this method leads to a very low relative estimated error in the magnitude of $10^{-6}$, but given the low complexity of the method, choosing a relatively large value $N$ presents no significant drawbacks. For the Fourier series method, the value $N=100$ is still deemed to give an acceptably low relative estimated error in the magnitude of $10^{-2}$, while any larger value would be too computationally expensive. 

\bibliography{ref} 

\end{document}